\documentstyle[preprint,aps]{revtex}
\begin{document}
\preprint{Usach 950301, hep-th 9503xxx}
\def\D{{\cal D}}
\def\12{{1\over 2}}
\def\st{\displaystyle}
\title{Anyons in $1+1$ Dimensions}
\author{J. Gamboa$^1$\thanks{E-mail: jgamboa@lauca.usach.cl} and
J. Zanelli$^{1,2}$\thanks{E-mail: jz@cecs.cl, jz@lauca.usach.cl} \\
$^1$ Departamento de F\'{\i}sica, Universidad de Santiago de Chile,\\
Casilla 307, Santiago, Chile\\ $^2$ CECS, Casilla 16443, Santiago, Chile}
\maketitle

\begin{abstract}

The possibility of excitations with fractional spin and statististics in
$1+1$ dimensions is explored. The configuration space of a two-particle
system is the half-line. This makes the Hamiltonian self-adjoint
for a family of boundary conditions parametrized by one real number
$\gamma$. The limit $\gamma \rightarrow 0, (\infty$) reproduces the
propagator of non-relativistic particles whose wavefunctions are even
(odd) under particle exchange. A relativistic ansatz is also
proposed which reproduces the correct Polyakov spin factor for the
spinning particle in $1+1$ dimensions. These checks support validity
of the interpretation of $\gamma$ as a parameter related to the
``spin'' that interpolates continuously between bosons ($\gamma
=0$) and fermions ($\gamma =\infty$). Our approach can thus be useful
for obtaining the propagator for one-dimensional anyons.

\end{abstract}
 \narrowtext

{\bf 1. Introduction}

Physics in $2+1$ dimensions has been extensively studied in the past few
years in connection with topological field theory \cite{witten} and with
the hope of finding interesting applications in condensed matter systems
\cite{fradkin}. In this last context, the interest arises from
experimental observations such as the quantum Hall effect or high-$T_c$
superconductivity in systems that are intrisically two-dimensional and
whose quantum excitations are expected to be anyonic.

A system of two particles in a plane is the manifold  $I\!R^2 - \{ 0
\}$ and for this reason the configuration space of a system of many
particles in 2+1 dimensions is multiply connected. In 2+1 dimensions
a particle interacts with another as with a composite object made out
of a particle and a magnetic flux. The dynamics of a system of particles
in the plane can thus be understood via an Aharonov-Bohm mechanism and
the fractional statistics can be seen to emerge as a consequence of this
fact. Furthermore, in 2+1 dimensions, the rotation group is abelian and
therefore the angular momentum need not be quantized \cite{leinaas}.

In $1+1$ dimensions the situation is topologically quite different and the
existence of unusual statistics is hardly mentioned. A common procedure
aimed at extracting information about the dynamics of (1+1)-dimensional
systems is bosonization {\it i.e.}, a transformation that expresses the
fermions in terms of bosonic fields and vice-versa. It is tacitly assumed,
however, that the quantum excitations can be either bosonic or fermionic,
and nothing else. Nevertheless, a minor modification of the usual
Bose-Fermi transformations can lead to fields with intermediate statistics.

Furthermore, in one spatial dimension the rotation group is discrete and
abelian so that, strictly speaking, spin does not exist. \footnote{ The
rotation group in 1 spatial dimension reduces to the discrete group of
reflections on a plane, {\bf Z$_2$}, so that the notion of angular momentum
as related to the dimension of its irreducible representations becomes
meaningless.} In complete analogy with the (2+1)-dimensional case, one
could also expect a continuous interpolation of spin and statistics
between bosons and fermions because the configuration space for a system
of particles is also multiply connected and the group of reflections is
abelian as well.

The possibility of having unusual spin and statistics in 1+1 dimensions is
theoretically very attractive also because there exist one-dimensional
systems (quantum wires) where properties like fractional spin could
perhaps be experimentally observed\cite{fischer}.

Although the continuous interpolation of spin and statistics in $1+1$
dimensions was suggested by Leinaas and Myrheim \cite{leinaas} (see also
\cite{anezeris} and \cite{poli}), the discussion found in the literature
still remains rather formal. (Recently, the possibility of fractional
statistics in $1+1$ dimensions has been considered following the
definition of statistics proposed by Haldane\cite{haldane}, \cite{papers}.
In this article we shall not adopt that approach.)

In (2+1) dimensions the effective theory obtained after integrating the
Chern-Simons field yields an explicit relation between the spin and the
Chern-Simons coefficients. There is no analogous construction in
(1+1) dimensions because there is no Chern-Simons action in this case.

The purpose of this note is to present a 1+1 dimensional model with the
properties mentioned above and to show how the notion of spin can be
related to the physical parameters of the interaction.

{\bf 2. Non-relativistic Model}

Let us consider two non-relativistic identical particles moving on a line.
For this system the configuration space is multiply connected (the real
line minus the origin) because the point where the particles collide is
singular. Classically the particles cannot go through each other, bouncing
off elastically every time they meet. Thus, the action of this system is
defined on the half-line [3,5], i.e.

\begin{equation}
S~=~ \int^{t_2}_{t_1} dt\,{1\over 2} {\dot x}^2,
\,\,\,\,\,\,\,\,\,\,\,\,\,\,\,\,\,\,\, (0<X< \infty),
\label{lagrang}
\end{equation}
where $x$ is the relative position of the two particles. As it is well
known, the Hamiltonian associated to (\ref{lagrang}) is not self-adjoint
on the naive Hilbert space because there is no conservation of
probability at $x=0$. The Hamiltonian for (\ref{lagrang}), however, can
be made self-adjoint by adopting a class of boundary conditions for all
the states in the Hilbert space of the form \cite{von}\cite{simon}

\begin{equation}
\psi^{'} (0) = \gamma \psi (0),
\label{condition}
\end{equation}
where $\gamma$ is an arbitrary real parameter.

The calculation of the propagator to go from an initial relative position
$x_1$ to a final one $x_2$ for the above problem gives \cite{sharp},
\cite{fahri}

\begin{equation}
G_\gamma [x(t_2), x(t_1)] = G_0 (x_2 - x_1) + G_0 (x_2 + x_1)
- 2\gamma \int_0^\infty d\lambda e^{-\gamma \lambda} G_0 (x_2 + x_1 + \lambda),
\label{full}
\end{equation}
where $G_0$ is the Green function for a free non-relativistic particle,
{\it i.e}

\begin{equation}
G_0 (x - y) = {1\over \sqrt{2\pi i t}} {\st e}^{i{(x - y)}^2/ 2t}. \label{free}
\end{equation}

Although in one spatial dimension it is  not possible to rotate particles,
they can be exchanged and their \lq \lq spin" and
statistics can be determined by the (anti-) symmetry of the wave
function. This (anti-) symmetry in turn depends on the values of the
parameter $\gamma$.

This last fact can be seen by taking the limits $\gamma = 0$
and $\gamma = \infty$ of (3) \cite{fahri}

\begin{equation}
G_{\gamma= 0, \infty} = G_0 (x_2 - x_1) \pm G_0(x_2 + x_1),
\label{b-f}
\end{equation}

Under exchange of the positions of two particles in initial or final states,
$G_{\gamma=0}$ is even and $G_{\gamma =\infty}$ is odd. Thus, for
$\gamma = 0$ ($\gamma = \infty$) and the particles behave as bosons
(fermions). The cases $0 < \gamma <\infty$ give particles with fractional
spin and statistics \cite{leinaas}.

The propagator (\ref{full}) can also be obtained in the path integral
representation, summing over all paths $-\infty < x(t) < \infty$, but in the
presence of a repulsive potential $\gamma \delta(x)$. This problem was
considered in\cite{sharp} -\cite{fahri} and the result is

\begin{equation}
G_\gamma [x(t_2), x(t_1)] ~
=~ \int {\cal D} x(t) \,\,e^{iS},
\label{prop1}
\end{equation}
with
\begin{equation} S~=~ \int_{t_1}^{t_2} dt ~\bigg(\12{\dot x}^2 + \gamma
\delta (x(t)) \biggr),
\label{act1}
\end{equation}
Here $\D x(t)$ is the usual functional measure. The potential term
$\gamma~ \delta(x(t))$ can be interpreted as a semi-transparent
barrier at $x=0$ that allows the possibility of tunneling to the
other side of the barrier. This is just another way of expressing
the possibility of interchanging the (identical) particles.

It is also interesting to note here that although in (1+1) dimensions
the rotation group is discrete and the definition of the spin is a
matter of convention, one may nevertheless view the one-dimensional
motion on the half-line as a radial motion with orbital angular
momentum $l=0$ \cite{grosche} in a central potential. This gives
rise to another possible definition of spin by taking the following
representation for the $\delta$-function

\begin{equation} \delta (x) = \lim_{\epsilon \rightarrow 0}
{\sqrt{\epsilon}\over {x^2 + \epsilon}}.
\end{equation}

Making a series expansion around $\epsilon =0$, the leading term
$\gamma\sqrt{\epsilon}/x^2$ is analogous to the centrifugal potential for the
radial equation in a spherically symmetric system, with
$\sqrt{\epsilon}\gamma$ playing the role of an intrinsic angular momentum
squared. Thus the spin of the system ($s$) can be defined by

\begin{equation}
s^2 = \sqrt{\epsilon}\gamma.
\label{5}
\end{equation}

For real $s$ (\ref{5}) only makes sense when $\gamma>0$. This definition
is consistent with the bosonic limit $\gamma =0$. For the fermionic case,
the limit $\gamma = \infty$ mentioned above is to be interpreted as
simultaneous with the limit $\epsilon  \rightarrow 0$ so that
$\sqrt{\epsilon} \gamma =1/4$. It is in this sense that the non-relativistic
quantum mechanics on the half-line describes one-dimensional anyons. However,
the normalization $s=1/2$ for fermions is conventional.

{\bf 3. Relativistic Model}

In this section we generalize the previous results to the relativistic
case. In order to do this one may either canonically quantize two
free relativistic particles moving on the line, or use path integrals
methods.

Let us start discussing the first approach. In analogy with the
non relativistic case, one may write the relativistic generalization
of (1) as
\begin{equation}
S~=~ \int_{\tau_a}^{\tau_b} d\tau \biggl[ {1\over 2N}{\dot X}^2 -
{m^2\over 2} N  \biggr],
\label{act3}
\end{equation}
where $N$ is the einbein along the worldline, $X^{\mu} = X_2^{\mu}
-X_1^{\mu}$ is a spacelike four vector whose spatial component is
restricted to be positive. In (\ref{act3}) the \lq \lq conformal gauge"
$\tau_1=\tau_2=\tau$ has been chosen.

The next step is to construct a covariant generalization of (2),
namely
\begin{equation}
\partial_\mu \psi = \gamma_\mu \psi,
\label{nbc}
\end{equation}
where $\gamma_\mu = (\gamma_0, \gamma_1)$ is a two-vector whose
components gives the two parameters that make the operator
$\partial^2_0 - \partial^2_1 + m^2$ self-adjoint, and $\psi$
is a solution of the Klein-Gordon equation
\begin{equation}
( \partial^2_0 - \partial^2_1 + m^2 ) \psi = 0.
\label{kg}
\end{equation}

Conditions (11) are consistent only for spacelike $\gamma_\mu$. This
is because for timelike $\gamma$ there exists a reference frame in
which the particle would ``bounce" at $X^0 = 0$ reversing its
time direction and violating energy conservation. Thus, one concludes
that in an appropriate (``rest'') frame, $\gamma_\mu = (0, \gamma)$,
or equivalently,

\begin{eqnarray}
{\partial \psi\over \partial t} \vert_{t=0} &=& 0, \nonumber
\\
{\partial \psi\over \partial x} \vert_{x=0} &=& \gamma \psi (x=0),
\label{cbc}
\end{eqnarray}
are consistent boundary conditions.

The calculation of the relativistic propagator is straightforward.
In the proper-time gauge ${\dot N} = 0$, it gives
\begin{equation}
G [X(\tau_b), X(\tau_a)] =
\int_0^\infty dT \int_{-\infty}^\infty
dp_0 dp_1 e^{-i{T\over 2}(p^2_0 - p^2_1 + m^2)} \psi^\ast (X(\tau_b))
\psi (X(\tau_a)).
\label{propa1}
\end{equation}
Here $\psi$ are solutions of (\ref{kg}) with the boundary condition
(\ref{nbc}), namely
\begin{equation}
\psi = e^{ip_0 X^0} \biggl[ e^{ip_1 X^1} + R e^{-ip_1 X^1} \biggr],
\label{solution}
\end{equation}
where $R$ is the reflection coefficient defined as
\begin{equation}
R = {{ip_1 -\gamma}\over {ip_1 + \gamma}}.
\end{equation}

Inserting (\ref{solution}) in (\ref{propa1}) one finds the
relativistic generalization of (3):

\begin{eqnarray}
G_\gamma [ X(\tau_b), X(\tau_a)] &=& {\cal N} {\st
\int}_0^\infty dT T^{-1} \exp \biggl[ i{{(\Delta X^0)}^2\over
2T} - i{m^2\over 2}T \biggr] \biggl( \exp \biggl[-i{{(X^1_b
-X^1_a)}^2\over 2T}\biggr] + \nonumber \\& & exp \biggl[-i{{(X^1_b
+ X^1_a)}^2\over 2T}\biggr]\biggr) - 2\gamma F_\gamma,
\label{propag}
\end{eqnarray}
where ${\cal N}$ is a normalization constant, $\Delta X^=X^0_b -X^0_a $
\begin{eqnarray}
F_\gamma &=& \int_0^\infty d\lambda
\int_0^\infty  dT T^{-1}e^{i(\Delta X_0)^2/2T - im^2T/2 -\gamma \lambda +
i(X^1_a + X^1_b + \lambda)^2/2T} \nonumber
\\
&=& \int_0^\infty d \lambda e^{-\gamma \lambda}
G_0 [X(\tau_b) + X(\tau_a)+ \lambda]
\end{eqnarray}
and $G_0 [X(\tau_b) - X(\tau_a)]$ is the free relativistic propagator,

\begin{equation}
G_0 [X(\tau_b) - X(\tau_a)] = \int_0^\infty dT T^{-1}e^{i(\Delta X)^2/2T
-im^2T/2}.
\end{equation}

Now let us consider the path integral formulation. The generalization
of (7) is the action\footnote{For simplicity we work
here in Euclidean space}
\begin{equation}
S~=~{1\over 2} \int_{\tau_a}^{\tau_b} d\tau \biggl[ {1\over N}{\dot X}^2 +
m^2 N + NV\biggr],
\label{actionn}
\end{equation}
where $V$ is a contact potential chosen so that in an instantaneous
rest frame the singularity occurs along the spatial direction,
\begin{equation}
V|_{rest}=\gamma \delta(X^1 (\tau)).
\end{equation}

In a generic frame the contact term would have a different expression
but its physical meaning should remain unchanged. The presence of a
$\delta$ function in the action allows the inclusion in the sum over
histories of trajectories that extend to $X^1<0$. These paths can be
taken into account if for each one that bounces off at $X=0$ one also
includes the path produced by the reflection of the bounce to $X^1<0$
\cite{fahri}. This means changing the final state $X_b$ according
to $X^1_b \rightarrow -X^1_b$, which is just an exchange of the
particles in the final state.

Taking into account the contact terms, the propagator of the
relativistic particle on the half-line in the proper time gauge
($\dot N =0$), is obtained:

\begin{eqnarray}
G_\gamma [ X(\tau_b), X(\tau_a)] &=& {\st \int}_0^\infty dN(0) \exp
\biggl[-{m^2 N(0)\Delta t\over 2}\biggr]
{\st \int} \D X^0 \exp \biggl[-{\st \int}_{\tau_a}^{\tau_b}d \tau {1\over
2N(0)} ({\dot X}^0)^{2}\biggr] \\ \nonumber
 & \times & {\st \int} \D X^1 \exp \biggl[-\st \int_{\tau_a}^{\tau_b}d \tau~
\biggl({1\over 2N(0)} ({\dot X}^1)^2 +i \gamma \delta (X^1(\tau))
\biggr) \biggr],
\end{eqnarray}
where $N(0)$ is the zero mode of $N(\tau)$.

The path integral over $X^0$ can be computed in analogy with a
non-relativistic particle with \lq mass\rq~ $N^{-1}(0)$ in the interval
$-\infty <X^0<+\infty$.  The result is

\begin{equation}
\int \D X^0 \exp \biggl[-\st \int_{\tau_a}^{\tau_b}d \tau~ {(\dot X^0)^2
\over 2N(0)}\biggr] =  ~\frac{1}{\sqrt{2\pi iT}} \exp \biggl[-{(\Delta
X^0)^2\over 2T}\biggr] \end{equation} with $T= N(0)\Delta \tau$.

The integral in $X^1$ can now be computed as the path integral for a
non-relativistic particle with \lq mass\rq~ $N^{-1} (0)$ moving on the
interval $-\infty <X^1 < \infty$. The relativistic propagator is then
\begin{eqnarray}
G_\gamma [X(\tau_b), X(\tau_a)] &=& G_0 [X(\tau_b) - X(\tau_a)] +
G_0 [X(\tau_b) + X(\tau_a)] \nonumber
\\
&-& 2\gamma \int_0^\infty d\lambda e^{-\gamma \lambda}
G_0 [X(\tau_b) + X(\tau_a)+ \lambda].
\label{full'}
\end{eqnarray}

One may now check the consistency if (\ref{full'}) by taking the
non-relativistic limit of (\ref{propag}). The integration in $T$ is
straightforward and the result is

\begin{eqnarray}
G_{\gamma} [ X(t_b), X(t_a)]~=& {\cal N} \biggl[ K_0 \biggl(
mc \Delta t \sqrt{1 -{{(X^1_b - X^1_a)}^2\over c^2 {\Delta t}^2}} \biggr) +
K_0 \biggl(
mc \Delta t \sqrt{1 -{{(X^1_b + X^1_a)}^2\over c^2 {\Delta t}^2}} \biggr)
 \\ \nonumber &
- 2 \gamma {\st \int}_0^\infty d \lambda e^{-i\gamma \lambda}
K_0\biggl(mc\Delta t \sqrt{1 + {{(X^1_b + X^1_a + \lambda)}^2 \over 2c^2
{\Delta t}^2}}\biggr)\biggr],
\label{nrl}
\end{eqnarray}
where we have inserted explicitly the speed of light ($c$) and ${\cal
N}$ is a normalization constant. In the limit $c\rightarrow \infty$
the Bessel
functions $K_\nu$ become

\begin{equation}
K_\nu (z) \rightarrow \sqrt{{\pi\over z}}e^{-z},
\,\,\,\,\,\,\,\,\,\,\,\,\, z>>1,
\label{k}
\end{equation}
for any order $\nu$. Using (\ref{k}) in Minkowski space, (\ref{nrl})
becomes\footnote{ The exponential $e^{-mc^2\Delta t}$ (or $e^{imc^2\Delta
t}$, when rotated back to Minkowski space) can be absorbed in a
renormalization of the energy by $H\rightarrow H-mc^2$ or by a constant
shift in the Lagrangian $L\rightarrow L+mc^2$.}

\begin{eqnarray} G_{\gamma } [ X(t_b), X(t_a)]~=& G_0 (X_b - X_a) + G_0
(X_b + X_a) \\ \nonumber & - 2\gamma {\st \int}_0^\infty d\lambda e^{ -\gamma
\lambda + imc^2 \Delta t} G_0 (X_b + X_a + \lambda),
\end{eqnarray}
which is the correct integral representation for the non-relativistic
propagator on the half-line given in eq. (3).

Finally, one can observe that, after eliminating the Lagrange
multiplier $N(\tau)$ from the relativistic action --by solving its
own equation of motion--, the effective action for the system of two
particles in a line can be written as
\begin{equation}
S[X(\tau_2),X(\tau_1)] =m\int^{\tau_2}_{\tau_1} d\tau \sqrt{(\dot X)^2} +
\gamma \times n,
\label{efact}
\end{equation}
where $n$ is the number of times $X^1(\tau)$ vanishes in the interval
$\tau_1 <\tau <\tau_2$. In other words, the contact term (21) merely
counts the number of crossings at $X^1 =0$ of each trajectory.

{\bf 4. Conclusions}

{\bf a.} We have constructed a model for anyons in $1+1$ dimensions
that is the analog of the ($2+1$)-dimensional case, both in the
non-relativistic limit, as well as in its relativistic extension. The
model is based on the observation that classically a system of two
particles on a line is equivalent to that of one particle on a
half-line. The statistics of the system can be read off from the
propagator (3) in the non-relativistic case, and from (17) (or (26))
in the relativistic case, which is also the exact formula for a free
scalar field theory on the half-line. The model can be considered
also as the relativistic generalization of the problem discussed by
Clark, Menikoff and Sharp \cite{sharp} and Farhi and Gutmann
\cite{fahri}.

{\bf b.} The parameter $\gamma$, which is related to the properties of the
contact interaction between identical particles, is the analogue of the
Chern-Simons coefficient $\sigma$. The ($2+1$)-dimensional expression
for the spin, $s = \sigma/4\pi$, is replaced in $1+1$ dimensions by $s =
\sqrt{\epsilon} \gamma$. In the non-relativistic case, our model provides
a new interpretion of the model discussed in [12,13]. The
interpretation of $\gamma>0$ as the spin of the particle remains
valid in the relativistic case because (9) is the right
expression for the coefficient of the centrifugal potential for a
relativistic particle in a spherically symmetric field.

{\bf c.} The non-relativistic quantum system on the half-line is a
particular case of the Yang-Yang model for two particles \cite{yang},
and the relativistic result presented here can be seen as the
corresponding generalization of that model. This may also be seen as
the reason {\it in profundis} for the appearence of anyons in $1+1$
dimensions.

{\bf d.} The relativistic propagator in $1+1$ dimensions can also be
obtained for the special case of spin $1/2$ by computing the Polyakov spin
factor of the spinning particle. This calculation is straightforward
\cite{polyakov} and gives the effective action $m\sqrt{{\dot X}^2} + \nu$,
where $\nu$ is the number of self-intersections of the wordline
\cite{strominger}.

Our results seem to agree in spirit with the spin factor calculation
of \cite{polyakov}, but with $\nu=\gamma n$ [c.f. eq. (\ref{efact})],
where $n$ is the number of collisions between the particles [e.g.,
collisions with the barrier at $X^1=0$, in the formulation of the
problem with the restriction $X^1>0$, or the number of crossings
through the barrier at $X^1=0$ in the formulation without that
restriction (using the contact potential)]. The connection with the
spin factor calculation can be made more manifest by the following
argument: Take all the nonintersecting paths that join a given
pair of points in $X^1$--space, and for each of them, take also its
reflection about $X^1=0$. Then the number of self intersections of the loop
formed by the two paths (joining their ends at $t=-\infty$ and at
$t=+\infty$), is just $n$. So, modulo a normalization, the effective
action obtained by Polyakov's approach is the same as the one given
here. There remains a puzzling point, however, because the two
actions would exactly match for $\gamma=1$, which is neither the
bosonic ($\gamma=0$) nor the fermionic case ($\gamma=\infty$).

\newpage

It is a pleasure to acknowledge many helpful comments and suggestions
by Professor Roberto Iengo, who was involved in the initial stages
of this work. We would like to thank also Professor Abdus Salam,
Unesco and the International Atomic Energy Agency for hospitality at
the ICTP, Trieste. This work was partially supported by grants
193.0910, 194.0203 and 195.0278 from FONDECYT (Chile), by
DICYT-USACH, by a European Communities research contract, by
institutional support to the Centro de Estudios Cien\-t{\'{\i}}\-fi\-cos de
Santiago provided by SAREC (Sweden), and by a group of chilean
private companies (COPEC, CMPC, ENERSIS, CGE). This research was also
sponsored by CAP, IBM and XEROX de Chile. J.G also thanks the Alexander
von Humboldt Foundation for financial support.

\end{document}